\documentclass[11pt]{article}
 \usepackage{hyperref}

\begin{document}

\baselineskip=17pt

\renewcommand{\abstractname}{\hfill}

%%%% *****************************************************************
%%%% *************    Text stat'i     ********************************
%%%% *****************************************************************
\newpage
\pagenumbering{arabic}
\thispagestyle{plain}

\begin{center}
\LARGE \bf Back reaction of the gravitational radiation\\
on the metric of spacetime
\end{center}

\begin{center}
{\bf A. A. Grib}\\[4pt]
{\sl A. Friedmann Laboratory for Theoretical Physics,\\
St.\,Petersburg, Russia\\[5pt]
Theoretical Physics and Astronomy Department, The Herzen  University,\\
48 Moika, St.\,Petersburg, 191186, Russia\\
andrei\_grib@mail.ru}\\[11pt]
{\bf Yu. V. Pavlov}\\[4pt]
{\sl Institute of Problems in Mechanical Engineering of
Russian Academy of Sciences,\\
61 Bolshoy, V.O., St.\,Petersburg, 199178, Russia\\[5pt]
N.I.\,Lobachevsky Institute of Mathematics and Mechanics,\\
Kazan Federal University, Kazan, Russia\\
yuri.pavlov@mail.ru}
\end{center}

\begin{abstract}
    The problem of back reaction of the gravitational radiation of
the two merging black holes on the metric of the space-time is investigated.
    It is shown for some models that large energy density of
the gravitational waves close to the merger can lead to the disappearance of
the visible accretion disc of the merged pair of black holes.
\noindent
PACS Number(s): {04.30.Tv, 04.70.Bw, 97.10.Gz}
\end{abstract}

\markboth{A. A. Grib \& Yu. V. Pavlov}
{Back reaction of the gravitational radiation on the
metric of spacetime}

%% pacs{04.30.Tv}{Gravitational-wave astrophysics}
%% pacs{04.70.Bw}{Classical black holes}
%% pacs{97.10.Gz}{Accretion and accretion disks}

%%%% *****************************************************************
\section{Introduction} \label{sec:intro}

    At present one has five well based events of the observation of
the gravitational
waves~\cite{GravWave16}--\cite{GravWave17c}.
    Majority of them are interpreted as the results of the merging black holes
located at the cosmological distances from the Earth.
    Numerical relativity was used in order to obtain the observable gravitational
waves as solution of nonlinear Einstein's
equations~\cite{Pretorius05}--\cite{Baker06}.
   The data of these observations are summed in the following Table~\ref{Table 1},
where $G$ is the gravitational constant, $c$ is the light speed.
\begin{table}[ht]
\arrayrulewidth=0.25mm
\begin{center}
\caption{Properties of the sources of the observed gravitational waves.}
\vspace{7pt}
%%%%%%%%%%%%%%%%%%%%%%%%%%%%%%%%%%%%%%%%%%%%%%%%%%%%%%%%%%%%%%%%%%%%%%%%%%%%%%%%%%%%%%%
{\begin{tabular}{|l|c|c|c|c|}  \hline
& GW150914 & GW151226 & GW170104 & GW170814
\\  \hline
Primary black hole mass $\frac{M^{\mathstrut}_1}{M_\odot} $
& $ 36^{+5}_{-4} $ & $ 14.2^{+8.3}_{-3.7} $ & $31.2^{+8.4}_{-6.0}$
& $30.5^{+5.7}_{-3.0}$
\\ \hline
Secondary black hole mass $ \frac{M^{\mathstrut}_2}{M_\odot} $
& $29^{+4}_{-4}$ & $7.5^{+2.3}_{-2.3} $ & $19.4^{+5.3}_{-5.9}$
& $25.3^{+2.8}_{-4.2}$
\\ \hline
Final black hole mass $\frac{M^{\mathstrut}_{f}}{M_\odot}$
& $62^{+4}_{-4}$ & $20.8^{+6.1}_{-1.7}$ & $48.7^{+5.7}_{-4.6}$
& $53.2^{+3.2}_{-2.5}$
\\ \hline
Radiated energy ($M^{\mathstrut}_\odot c^{2} $)
& $3.0^{\mathstrut +0.5}_{-0.5}$ & $1.0^{+0.1}_{-0.2}$ & $2.0^{+0.6}_{-0.7}$
& $2.7^{+0.4}_{-0.3}$
\\ \hline
Peak luminosity ($10^{\mathstrut 56}$\,erg\,s$^{-1}$)
& $3.5^{\mathstrut +0.5}_{-0.4} $ & $3.3^{+0.8}_{-1.6}$ & $3.1^{+0.7}_{-1.3}$
& $3.7^{+0.5}_{-0.5}$
\\ \hline
Frequency $\nu$ (Hz)
& $35 - 250 $ & $35 - 450$ & $35 - 400 $ & $ 40 - 250 $
\\ \hline
Luminosity time $\Delta t $ (s) $\approx$
& $ 0.1$ & $ 1$ & $ 0.1$ & $ 0.1 $
\\ \hline
Final black hole spin $A = \frac{c|J|^{\mathstrut}}{G M_{f \mathstrut}^{2}}$
& $0.67^{+0.05}_{-0.07} $ & $0.74^{+0.06}_{-0.06}$ & $0.64^{+0.09}_{-0.20}$
& $0.70^{+0.07}_{-0.05}$
\\  \hline
\end{tabular}  \label{Table 1} }
\end{center}
\end{table}
    One can see that in all cases of observation a huge energy of the order of
some number of Solar masses $M_\odot c^2$ is radiated at a short time
$\Delta t\approx 0.1 - 1$\,s
in small region of the space of the order of $\sim (c/\nu)^3 $.
    The originating energy of radiation density is comparable only to
the energy density in the fireball of the early Universe in Big Bang.
    However the difference of the ``gravitational fireball'' due to
gravitational radiation from the electromagnetic radiation fireball is that
gravitational radiation very weakly interacts with matter and looks invisible
for it.
    Does it mean that it does not influence matter surrounding the merger?
        This problem was not investigated in
Refs.~\cite{Pretorius05}--\cite{Baker06}.

    Here we show that there is indirect influence of gravitational radiation
on matter due to the influence on the curvature of the space-time.
    Ricci term in Einstein equations can be considered to be
non zero due to gravitational radiation and particles of matter surrounding
the merged black holes move differently from the case when it was absent.

    It is well known that in cosmology of the early Universe such large energy
density changing in time leads to  the Friedmann model of the expanding
Universe.
    Can one suppose that back reaction of the energy density of gravitational
radiation close to the merged two black holes will lead to large expansion
of space in that region?

    This hypothesis means that the averaged in space effective stress
energy of the gravitational radiation of merged black holes,
when the dimensionless parameters $h_{ik}$ (see Sec.~35 in Ref.~\cite{MTW})
cannot be considered to be much smaller than 1,
is not much different from that of ultrarelativistic gas of
massless particles in homogeneous isotropic Friedmann space-time in cosmology.
    The exact form of the stress-energy terms in case where we cannot neglect
nonlinear effects is not known.
    However some estimates made by Isaacson~\cite{Isaacson68b}
for large frequencies of gravitational waves show that it is not much different
from the electromagnetic case.

    The bubble of gravitational radiation is expanding in space-time up to
some time when the density of gravitational radiation will be equal to
the density matter in the space surrounding the two merged black holes.
    In the region of space close to the merging black holes an accretion disc
could be formed.
    Due to more or less large time of evolution up to the merged pair
one cannot suppose that the accretion disc has large density of matter,
so particles can be considered as probe particles.
    That is why large energy density of gravitational radiation is propagating
in practically empty space up to some moment when our supposition about small
influence of the surrounding matter becomes incorrect.
    This  moment can be taken as that when  the gravitational energy radiation
becomes equal to the density of matter in usual conditions taken as
$\approx 1$\,g/cm$^3$.
    So suppose that quasi Friedmannian stage of expansion takes place in local
region of space  surrounding merging black holes.
    On large distances one has only gravitational wave predicted
by Refs.~\cite{Pretorius05}--\cite{Baker06}.
    To what observational consequences this hypothesis will lead?

    It is evident that one must answer on the question: how the expansion
stage will influence on the existence of the accretion disc?
    Energy is not conserved in nonstationary metric which leads to the
nonstability of orbits around the merging black holes.
    After the gravitational radiation is finished metric becomes stationary
and the energy will be conserved so that new stable orbits and accretion disc
will be formed.
    The details of the formation of the new disc depend on the distribution
of matter around the black hole.
    If it will disappear due to expansion then the electromagnetic
R\"{o}ntgen radiation from the merger will also disappear.
    The far observer will see sudden disappearance of the point source of
R\"{o}ntgen radiation.
    Let us do some calculations.

%%%% *****************************************************************
\section{Evaluation of the energy density of gravitation radiation of the merger}
\label{sec2}

    Let the summed mass of the merged black holes is
$ M = \beta\, M_\odot$
($ \beta \approx 65$ for the event GW150914).
    The radiated energy is $\Delta M \, c^2$,
where $\Delta M = \alpha M$
($\alpha \approx 3/65$ for the event GW150914).
    The evaluation of the size of the radiation region $r = c/ (2 \pi \nu)$
shows that it is close to the gravitational radius
$ R_g = 2 G M /c^2 $: \
$ r = \gamma R_g $
($\gamma = 2 - 10 $ for GW150914).
    Then one can obtain the gravitational energy density as
    \begin{equation}    \label{p3}
\varepsilon = \frac{\Delta M c^2}{\frac{4}{3} \pi r^3} =
\frac{3}{32 \pi}\, \frac{\alpha}{\beta^2 \gamma^3}\, \frac{c^8}{G^3 M_\odot^2}.
\end{equation}
    Putting into~(\ref{p3}) the values of parameters for the event GW150914:
$\alpha = 0.05$, \ $\beta = 65$, \ $\gamma = 10$, one obtains
    \begin{equation}    \label{ope}
\varepsilon = 3 \cdot 10^{29}\, {\rm J}/{\rm m}^3 .
\end{equation}
    As it is seen from Table~\ref{Table 1} the similar evaluation can be made
for all other events GW151226, GW170104, and GW170814.

    The obtained energy density is much smaller than Planckian values
$\varepsilon_{\rm Pl} = m_{\rm Pl} c^2 / l_{\rm Pl}^3 = c^7 / (\hbar G^2)
\approx 4.7 \cdot 10^{113}$\,J /m$^3$,
    but corresponds to the energy density of the early Universe at
the first second from the Big Bang, i.e. to the era of nucleons formation.

%%%% ****************************************************************
\section{Local cosmological expansion in the vicinity of the merged
pair of black holes}
\label{secRK}

    Let us use the model of the homogeneous isotropic expansion in
the region of space surrounding the merged pair of black holes.
    In this region the energy $\Delta M \, c^2$ appears at the short
moment of time leading to the energy density~(\ref{ope}).
    Note that this model is not exact because the gravitational radiation
cannot be spherically symmetric.
    So in case of the gravitational radiation of double stars the intensity
of radiation averaged over a period is proportional to
$1 + 6 \cos^2 \theta + \cos^4 \theta$, where $\theta$ is a polar angle
(see Sec.~1.13 in Ref.~\cite{ZeldovichNovikovTTES}).
    Really one can say that maximal intensity of radiation in direction
$\theta =0$ is related to the intensity of radiation in direction
$\theta = \pi/2$ as 8:1.
    Further we also discuss the role of anisotropy in consideration after
formula~(\ref{pewf}).
    There are also some other arguments indicating similarity of effects in
gravitational waves and in cosmology~\cite{Possel17}.
    So let us consider that a solution of Einstein equations
    \begin{equation}    \label{p5}
R_{ik} - \frac{g_{ik}}{2} R = - \frac{8 \pi G}{c^4} T_{ik}
\end{equation}
    in the region in the vicinity of merged black holes the metric is supposed
to be that of homogeneous isotropic quasi-Euclidean space-time of the form
    \begin{equation}    \label{p6}
d s^2 = c^2 dt^2 - a^2(t)\, d \mathbf{r}^2 .
\end{equation}
    It is preferable to take the stress energy tensor of the gravitational
radiation formed by gravitational waves as it is for the radiation dominated
matter (the pressure $p = \varepsilon/3$).
    However the calculation can be made for the general case with equation of
state
    \begin{equation}    \label{pw}
p = w \varepsilon, \ \ \ \ w= {\rm const} , \ \ \ \ (1 \ge w \ge 0).
\end{equation}
    In this case the solution of the Einstein equations is
    \begin{equation}    \label{ptw}
a (t) = a_0 t^q , \ \ \ \ q = \frac{2}{3 (w +1)}, \ \ \ \
\varepsilon (t) = \frac{3 c^2 q^2}{8 \pi G t^2},
\end{equation}
    \begin{equation}    \label{pew}
\varepsilon (t) = \varepsilon (t_0)
\left( \frac{a (t)}{a (t_0)} \right)^{-3(1 +w)}.
\end{equation}
    Putting into~(\ref{ptw}) the value~(\ref{ope}) for the energy density
one obtains for the time approximately the value $0.1$\,s (for any $w$ from
the permitted interval) which corresponds to the time of intensive radiation
in merging of black holes (see Table~\ref{Table 1}).

    For the case of radiation dominated matter one has
    \begin{equation}    \label{p7}
a(t) = a_0 \sqrt{t}, \ \ \
t = \frac{c}{4} \sqrt{ \frac{3}{2 \pi G\, \varepsilon(t)} }.
\end{equation}
    From the condition that the size of the radiation region at this moment is
equal to $\gamma R_g$, for $a_0$ in the scale factor one obtains
    \begin{equation}    \label{mf}
a_0 = \frac{\gamma R_g}{ \sqrt{0.12\, {\rm s}}}
\approx \gamma\, 400\, \frac{\rm km}{s^{1/2}},
\end{equation}
    where numerical estimate is made for $M = 65 M_\odot$ for GW150914.

    Let us suppose that expansion stops when the  energy density of
the gravitational radiation becomes equal to the energy density of usual
matter (taken for example as the water density
$ \varepsilon_1 = 9 \cdot 10^{19}$\,J/m$^3$).
    Then from~(\ref{pew}) one obtains
    \begin{equation}    \label{pewf}
\frac{a (t)}{a (t_0)} =
\left( \frac{\varepsilon (t_0)}{\varepsilon (t)} \right)^{
1/\left[3(1+w)\right]}
\sim 10^{10/\left[3(1+w)\right]}.
\end{equation}
    So the expansion is equal to $\approx 2000 $ (for $w=0$),
$\approx 300 $ (for $w=1/3$), $\approx 50 $ (for the limiting case $w=1$).
    For all cases the expansion is large.

    From~(\ref{pewf}) one can see that if $\varepsilon (t_0)$ is different
in different directions as 8:1 then there is small difference in $a(t)$ in
these directions defined by $8^{1/\left[3(1+w)\right]} $
($8^{1/4}$ for $w=1/3$) which can be neglected.

    Let us show that this leads to disappearance of the visible accretion disc.

%%%% ****************************************************************
\section{The accretion disc and local expansion}
\label{seckor}

    The influence of the general cosmological expansion on the local
properties of gravitationally bound systems was widely discussed in
literature (see for example Refs.~\cite{FaraoniJ07,Price12}
and references there).

    Consider free movement of the particle on homogeneous isotropic
expanding space.
%%%%%%%%%%%%%%%%%%%%%%%%%%%%%%%%%%%%%%%%%%%%%%%%%%
    As it is known~\cite{Chandrasekhar},  equations of geodesics in
the space-time with the interval $ds^2 = g_{ik} dx^i dx^k$ can be obtained
from the Lagrangian
    \begin{equation}
L = \frac{g_{ik}}{2}\, \frac{ d x^i}{d \lambda} \frac{ d x^k}{d \lambda},
\label{Lgik}
\end{equation}
    where $\lambda$ is the affine parameter on the geodesic.
    For particle with nonzero rest mass $m$ the parameter $\lambda = \tau/m$,
where $\tau$ is the proper time of the massive particle.

    Generalized momenta are by definition
    \begin{equation}
p_i  \stackrel{\rm def}{=} \frac{\partial L}{\partial \dot{x}^i}
= g_{ik} \frac{d x^k}{d \lambda } ,
\label{Lpdef}
\end{equation}
    where $ \dot{x}^i = d x^i/d \lambda $.
    If the metric components $g_{ik}$ do not depend on some coordinate $x^n$
then the corresponding canonical momentum (the corresponding covariant component)
is conserved in motion along the geodesic due to Euler-Lagrange equations:
    \begin{equation}
\frac{d }{d \lambda} \frac{\partial L}{\partial \dot{x}^n} -
\frac{\partial L}{\partial x^n} = 0, \ \
\frac{\partial g_{ik}}{\partial x^n} = 0 \ \ \Rightarrow \ \
p_n = \frac{\partial L}{\partial \dot{x}^n} = {\rm const}.
\label{LEL}
\end{equation}
    The components of the 3-momentum of the particle are conserved in
the homogeneous isotropic expanding space with metric~(\ref{p6})
    \begin{equation}
p_\alpha = a^2(t) \frac{d x^\alpha}{d \lambda} =
 m a^2(t) \frac{d x^\alpha}{d \tau} = {\rm const} .
\label{imp}
\end{equation}
    If the expansion stopped ($a(t)={\rm const} = a_f$, for $t \ge t_f$),
then for time $t \ge t_f$ the values $ a_f x^\alpha$ were measurable physical
distances and $ d (a_f x^\alpha) / d \tau$ are velocities.
    From eq.~(\ref{imp}) it is seen, that after expanding in $k$-times,
the velocity becomes smaller also in $k$-times.

    One can obtain the same conclusion from conservation of the angular
momentum in homogeneous isotropic expanding space.
    So the projection of angular momentum of the particle on the rotation axis
$ J = m r v_\bot$ must be conserved.
    Here $r$ is distance to the rotation axis.
    That is why if the space expanded in $k$-times then the velocity~$v_\bot$
perpendicular to the radius becomes smaller in $k$-times.

    Let us apply these results to the accretion disc in expanding space around
merging black holes.
    If the trajectories of particles in the disc are close to circular
then the radius becomes $k$-times larger due to the expansion $r_f = k r_0$
and the velocity perpendicular to the radius becomes $k$-times
smaller $v_f = v_0/k$.
    This velocity occurs much smaller than that needed for rotation with
the new radius and particle will move along much prolonged trajectory around
the black hole.

    Let us show this first for movement of the particle in Newtonian
potential $U(r) = - G m M /r$.
    For circular rotation of the particle on the distance $r_0$ from the centre
one has the velocity of rotation $v_0 = \sqrt{\mathstrut G M/r_0}$,
the projection of the angular momentum $J = m \sqrt{\mathstrut G M r_0}$
and full energy
    \begin{equation}    \label{ra1}
E_0 = \frac{m v_0^2}{2} - G \frac{m M}{ r_0} = - G \frac{m M}{2 r_0}.
\end{equation}
    After expansion the full energy is
    \begin{equation}    \label{ra2}
E_f = \frac{m v_f^2}{2} - G \frac{m M}{ r_f} =
- G \frac{m M}{2 r_f} \left( 2 - \frac{1}{k} \right).
\end{equation}
    The boundaries of the region of possible movement are defined by
inequality~\cite{LL_I}
    \begin{equation}    \label{ra3}
E_f \ge U(r) + \frac{J^2}{2 m r^2} .
\end{equation}
    From~(\ref{ra3}) one sees that particle movement after expansion of space
in $k$-times will be on elliptical orbit with maximal distance $r_f$
and minimal distance $r_{\rm min} = r_0 / \left( 2 - \frac{1}{k} \right) $.
    For particles being initially at different points on the same orbit
these ellipses will be different.

    Now consider relativistic case of the Schwarzschild metric (neglecting
black hole angular momentum) of the merger:
    \begin{equation}
d s^2 = \biggl( 1 - \frac{2M}{r} \biggr) d t^2 -
\frac{ d r^2 }{\displaystyle 1 - \frac{2M}{r} }
- r^2 \! \left( \sin^2\! \theta \, d \varphi^2 + d \theta^2 \right).
\label{g1}
\end{equation}
    Here we use the system of units $G = c =1$.
    The equations for geodesics in metric~(\ref{g1}) can be written
for $\theta = \pi/2 $ as
    \begin{equation} \label{geodSch1}
\frac{d t}{d \lambda} = \frac{r}{r-2M}\, E,
\ \ \ \
\frac{d \varphi}{d \lambda} = \frac{J}{r^2},
\end{equation}
    \begin{equation} \label{geodSch3}
\left( \frac{d r}{d \lambda} \right)^2 =
E^2 + \frac{2 M - r}{r^3} J^2 + \frac{2 M - r}{r} m^2,
\end{equation}
    where $E$ is the energy of the moving particle,
$J$ --- the conserved projection of the particle angular momentum on the
axis orthogonal to the plane of movement,
$m$ is the mass of the test particle.
    Define the effective potential by the formula
    \begin{equation} \label{Leff}
V_{\rm eff} = -\frac{1}{2} \left[
E^2 + \frac{2 M - r}{r^3} J^2 + \frac{2 M - r}{r} m^2 \right].
\end{equation}
    Then
    \begin{equation} \label{LeffUR}
\frac{1}{2} \left( \frac{d r}{d \lambda} \right)^{\!2} + V_{\rm eff}=0, \ \ \ \
\frac{d^2 r}{d \lambda^2} = - \frac{d V_{\rm eff}}{d r}.
\end{equation}

    If the particle orbit in space around the black hole is bounded,
then in the point of maximal length one must have the following conditions
for the effective potential
    \begin{equation} \label{Leffusl}
V_{\rm eff}=0, \ \ \ \
\frac{d V_{\rm eff}}{d r} \ge 0.
\end{equation}
    In the bottom point of a trajectory one has
    \begin{equation} \label{Leffnt}
 V_{\rm eff}=0, \ \ \ \ \frac{d V_{\rm eff}}{d r} \le 0.
 \end{equation}
    For the circular movement  $ V_{\rm eff}=d V_{\rm eff}/d r=0$ and at
radial distance $r_0$ one has~\cite{ShapiroTeukolsky}
    \begin{equation}    \label{ek4}
\frac{J^2}{m^2} = \frac{M r_0^2}{r_0 - 1.5 R_g}.
\end{equation}
    Note that circular orbits for massive particles are possible only
for $r> 1.5 R_g$.
    The circular orbits are stable if the radial coordinate $r > 3 R_g$.
    Particle energy on the stable orbit $E < m$ and on the minimal stable
circular orbit is $E = 2 \sqrt{2} m /3$.

    Consider movement of particle in metric~(\ref{g1})
if due to expansion it was taken from the circular orbit on the distance $r_0$
to the radial distance $r_f = k r_0$ with conservation of the projection of
angular momentum $J$.
    The new value of the energy $E_f$ can be found from
the condition $ V_{\rm eff}(r_f)=0$:
    \begin{equation} \label{Efn}
E_f^2 = \frac{r_f - 2 M }{r_f} \left( \frac{J^2}{r_f^2} + m^2 \right).
\end{equation}
    One can find possible region of particle movement after space expansion
from the condition $ V_{\rm eff}(r) \le 0$ putting there the value $E_f^2$.
    For $r_0 > 3 R_g$ one obtains $r \le r_f$.
    If $r_0 > 6 R_g$ taking into account that the expansion is large $k \gg 1$
one obtains from~(\ref{Leffusl})--(\ref{ek4}) for
the possible interval $2 R_g < r_{\rm min} < r \le r_f$.
    So for Schwarzschild metric the new particle trajectories for the accretion
disc will be strongly prolonged orbits.

    The increase of the disc in $k$-times will result in its strong cooling.
    The temperature is proportional to the square of the average velocity of
relative particle movement.
    Due to fact that velocities become smaller in $k$-times the relative
velocities also become smaller in $k$-times.
    So the temperature become smaller in $k^2$-times.

    The same evaluation of the temperature change one obtains from
the adiabatic equation of an ideal gas
    \begin{equation}    \label{adia}
T V^{(\gamma -1)} = {\rm const},
\end{equation}
    where $T$ is temperature, $V$ is volume of gas,
$\gamma$ is adiabatic index.
    For ionized plasm $\gamma=5/3$ and growth of all distances in $k$-times
leads to the decrease of the temperature in $k^2$-times.

    In our model the expansion of space close to merging black holes due to
gravitational radiation will be from 50 to 2000 times.
    So the temperature will decrease in $10^3 - 10^6$ times.
    The intensity of electromagnetic radiation due to Stefan-Boltzmann law
will decrease in $10^{12}$ times.
    So the accretion disc will become practically invisible.

    Then after the expansion will be finished the accretion disc can be again
compressed into a new compact radiating object.
    The details of this process depend on the mass and matter density of the
initial accretion disc.

%%%% ****************************************************************
\section{Conclusion} \label{secConcl}

    Now let us make some remarks on our investigation.
    Huge energy density of gravitational radiation propagating in space
creates due to Einstein equations the appearance of nonstationary metric for
space surrounding the source of this radiation.
    It is reasonable to think that this nonstationary metric somehow added to
the initial static metric of the merged pair of black holes leads to
the expansion of space.
    The accretion disc could be formed around the merged pair before
the flash of the gravitational radiation.
    The existence of the disc is due to the movement of probe particles on
some orbit which for simplicity can be considered to be circular.
    The appearance of the gravitational radiation from the merger due to
nonstationarity of the arising metric leads to the nonstability of the orbit.
    The calculations using simple models considered by us show that this
nonstability will be large enough to make the accretion disc invisible.
    We believe that this simple picture will survive in more complex case
when the nonstationary metric is not considered to be homogeneous and
isotropic in space.
    To observe the effect on the Earth the observer must look for events of
the sudden disappearance of some point sources of electromagnetic R\"{o}ntgen
radiation followed by the flash of gravitational radiation from these sources.

%%%% *****************************************************************
\section*{Acknowledgments}

    This work was supported by the Russian Foundation for Basic Research,
grant No. 18-02-00461-a and by the Russian Government Program of
Competitive Growth of Kazan Federal University.

%%%% *****************************************************************

\end{document}